\documentclass[aps,prl,reprint,superscriptaddress]{revtex4-2}

\usepackage{blindtext}
\usepackage{amsmath}
\usepackage{amsfonts}
\usepackage{graphicx}
\usepackage{appendix}
\usepackage{xcolor}
\usepackage{soul}

\graphicspath{ {./Figures/} }

\newcommand{\vres}{{v_\mathrm{res}}}
\newcommand{\vthr}{{v_\mathrm{thr}}}
\newcommand{\vmin}{{v_\mathrm{min}}}

\newcommand{\bra}[1]{\left\langle #1 \right|}

\newcommand{\braket}[2]{\left\langle #1 \right|\left. #2 \right\rangle}

\newcommand{\EqRef}[1]{Eq.~\eqref{#1}}
\newcommand{\FigRef}[1]{Fig.~\ref{#1}}

\begin{document}

% \title{The dynamical laws of neural populations: \\ relation between mean membrane potential and firing rate } % 0GVV
%\title{Nonperturbative moment dynamics of the membrane potential density in spiking neuron networks} % 1MM
% \title{Nonperturbative moment dynamics for networks of spiking neurons}% 2MM
\title{Self-consistent moment dynamics for networks of spiking neurons}% 2MM

% \author{Maurizio Mattia}
% \email{maurizio.mattia@iss.it}
% \affiliation{Natl. Center for Radiation Protection and Computational Physics, Istituto Superiore di Sanità, 00161 Rome, Italy}

% \author{Gianni V. Vinci}
%\email{giannivinci.42@gmail.com}
% \affiliation{Natl. Center for Radiation Protection and Computational Physics, Istituto Superiore di Sanità, 00169 Rome, Italy}
% \affiliation{PhD Program in Physics, “Tor Vergata” University of Rome, 00133 Roma, Italy}

\author{Gianni V. Vinci}
\affiliation{Natl. Center for Radiation Protection and Computational Physics, Istituto Superiore di Sanità, 00161 Roma, Italy}
\affiliation{PhD Program in Physics, “Tor Vergata” University of Rome, 00133 Roma, Italy}

\author{Roberto Benzi}
\affiliation{Dept. of Physics and INFN, “Tor Vergata” University of Rome, 00133 Roma, Italy}

\author{Maurizio Mattia}
\email{maurizio.mattia@iss.it}
\affiliation{Natl. Center for Radiation Protection and Computational Physics, Istituto Superiore di Sanità, 00161 Roma, Italy}

\date{March 5, 2025}

\begin{abstract}
A novel approach to moment closure problem is used to derive low dimensional laws for the dynamics of the moments of the membrane potential distribution in a population of spiking neurons. 
Using spectral expansion of the density equation we derive the recursive and nonlinear relation between the moments, such as the mean potential, and the population firing rates. 
The self-consistent dynamics found relies on the dominant eigenvalues of the evolution operator, tightly related to the moments of the single-neuron inter-spike interval distribution. 
Contrary to previous attempts our system can be applied both in noise- and drift-dominated regime, and both for weakly and strongly coupled population. 
We demonstrate the applicability of the theory for the case of a network of leaky integrate-and-fire neurons deriving closed analytical expressions. 
Truncating the mode decomposition to the first few more relevant moments, results to effectively describe the population dynamics both out-of-equilibrium and in response to strongly-varying inputs.
\end{abstract}

\maketitle

%\tableofcontents

%\paragraph{Middle-out approach to bridge micro- and macroscales ---}
Bridging microscopic neuron dynamics and macroscopic population behavior is challenging. 
Phenomenological models describe mesoscopic variables like mean membrane potential $m(t)$ or population firing rate $\nu(t)$ but lack mechanistic detail. 
Bottom-up mean-field models derive population behavior from single neurons but can be complex. 
A middle-out approach, focusing on the mesoscopic scale, combines mechanistic insight with tractable modeling, effectively linking micro- and macroscales \cite{Noble2002}.

% \paragraph{Low-dimensional dynamics from moment-closure approaches  ---}
Mean-field theories are often preferred because they link different scales, relating population firing rates to key single-neuron properties. 
However, closing these systems is challenging due to the nonlinear, out-of-equilibrium nature of neuronal dynamics, leading to approximations that limit their applicability. 
Early moment-closure approaches for neural networks  laid the groundwork for capturing population dynamics through truncated moment hierarchies \cite{Rodriguez1996, Tanabe2001}. 
More recently, elegant low-dimensional descriptions based on the Ott-Antonsen ansatz \cite{Ott2008} have been developed for networks of quadratic integrate-and-fire neuron \cite{Luke2013, Montbrio2015}, but these typically assume deterministic synaptic inputs -- an unrealistic condition for biological neurons. 
Advances in systematic and automated moment-closure methods \cite{Wuyts2022} offer promising tools to better handle stochasticity and network interactions.
Nonetheless, moment-closure approaches face fundamental challenges, particularly in deciding which moments to neglect or approximate \cite{Levermore1996}. 
There is indeed no clear, general criterion for selecting the appropriate closure, making the process somewhat heuristic and potentially limiting accuracy and stability. 

% \paragraph{Low-dimensional dynamics from spectral expansion ---}
Spectral decomposition methods  provides an alternative and general framework applicable to networks of arbitrary spiking neurons \cite{Abbott1993, Treves1993, Knight2000, Mattia2002}. 
However, this approach often struggles to fully capture coupling effects between neurons. 
Despite this limitation, spectral decomposition remains a powerful tool to reduce the complex, high-dimensional neuronal dynamics into a low-dimensional representation by extracting dominant modes of activity \cite{Schaffer2013, Mattia2016}. 
Developing a generalized framework that can accurately describe the rich repertoire of neuronal population dynamics without significant restrictions is crucial for guiding experiments and building a coherent understanding of the underlying physics of these systems.

% \paragraph{Our contribution ---}
In this \textit{Letter} we show a `third way' to derive a low-dimensional dynamics of the firing rate of a homogeneous network of spiking neurons. 
By relying on the first relevant modes of a spectral expansion, we compute the approximated moments of the probability density which in turn drive the coefficients of the expansion.
A self-consistent dynamics for the firing rate and the moments of such neuronal density is then unfold. 
The introduced macroscopic theory  can be straightforwardly applied to a wide class of spiking neuron models. 
Thanks to the moving basis adopted for the expansion \cite{Knight1996, Mattia2002}, its description power allows to reliably capture fast transients in response to exogenous perturbations and out-of-equilibrium regime of activity.

\paragraph{Population density approach ---}
Consider a network of interacting integrate-and-fire (IF) neurons. 
In the mean-field limit, where the number of neurons $N \to \infty$, the collective behavior of these spiking neurons is captured by the probability density $p(v, t)$, representing the likelihood of finding a neuron with membrane potential $v$ at time $t$ \cite{Abbott1993,Knight1996,Brunel1999,Fusi1999}. 
This density evolves according to the following Fokker-Planck (FP) equation
\begin{equation}
\partial_{t}p 
   = -\partial_{v}[(F+\mu) p] +\frac{1}{2} \partial^2_v (\sigma^2 p)
	\equiv \mathcal{L} \, p 
   \equiv -\partial_v S_{p} \, .
\label{eq:FP}
\end{equation}
In this continuity equation the density $p$ changes according to the divergence of the probability current $S_p(v,t) = (F+\mu) p - \frac{1}{2} \partial_v (\sigma^2 p)$.
Here, the membrane potential $V(t)$ of a neuron follows the Langevin equation
\begin{equation}
   dV = \left[F(V) + \mu\right]dt + \sigma dW \, ,
\label{eq:VDynDiff}
\end{equation}
where $F(V)$ is the drifting current determining the model-specific relaxation dynamics \cite{Gerstner2014}.
The synaptic current to neurons is modeled as Gaussian white noise $W(t)$ [$\langle W(t) W(t')\rangle=\delta(t-t')$] with time-dependent mean $\mu(V,t)$ and variance $\sigma^2(V,t)$. 
In cortical networks, this diffusion approximation is valid because the number $K$ of presynaptic contacts is large and synaptic efficacy $J$ is small \cite{DeFelipe2002,Markram2015}.
By the Grigelionis central limit theorem \cite{Grigelionis1963}, pooling independent, low-rate presynaptic spike trains leads to an inhomogeneous Poisson process. 
The assumption of small $J$ ensures independence, supporting the mean-field approximation, where all neurons are independent realizations of the same stochastic process (i.e., with same $F$, $\mu$ and $\sigma$) \cite{Amit1997}), drawn from the probability density $p(v,t)$.

Synaptic interactions are incorporated in the $\nu$-dependent moments of the current, which in current-based models are 
\begin{equation}
\begin{split}
	     \mu(\nu) & = K J \nu(t) + \mu_{ext}(t) \\ 
   \sigma^2(\nu) & = K J^2 \nu(t) + \sigma^2_{ext}(t)
\end{split} 
\label{eq:MuSigma}
\end{equation}
Here, the firing rate $\nu(t) = S_p(\vthr,t)$ is the flux of neurons crossing the emission threshold $\vthr$, and $\mu_{ext}$ and $\sigma^2_{ext}$ are the moments of the synaptic current due to the spikes incoming from  external populations. Note that, despite being one-dimensional, the gradient condition \cite{Risken1984} does not hold for \eqref{eq:FP} (i.e., the detailed balance is always broken) due to the peculiar boundary condition that follows from the reset mechanism of spiking neuron:
\begin{equation}
    \partial_v p(v_\mathrm{res}^+,t)- \partial_v p(v_\mathrm{res}^-,t)= \partial_v p(\vthr,t) \, ,
\label{eq:ReinjBC}
\end{equation}
where $\vres$ is the membrane potential at which the neurons are reset after the emission of a spike. 
Here, for the sake of simplicity, we assumed no absolute refractory period. 
Another important feature that complicates the analysis is that, due to the dependence of the synaptic current on $\nu$, the FP equation is also nonlinear \cite{Mattia2002}.

\paragraph{Dynamics of membrane potential moments ---}
The evolution in time of the moments of the potential $v$ 
\begin{equation}
   m_k(t) = \int_\vmin^\vthr v^k p(v,t) dv = \braket{v^k}{p}
\label{eq:VMoments}
\end{equation}
can be derived by applying the integral operator $\int_\vmin^\vthr v^k dv$ (i.e., the `bra' $\bra{v^k}$) to both hand sides of the FP \EqRef{eq:FP}:
\begin{equation}
    \dot{m}_k = \braket{v^k}{\partial_t p} = \braket{v^k}{\mathcal{L} p}.
\label{eq:MDynStart}
\end{equation}
This dynamics can be rewritten as a convenient series by resorting to the spectral decomposition
\begin{equation}
  \mathcal{L} \phi_n = \lambda_n \phi_n \, ,
\label{eq:SpecDecomp}
\end{equation}
where if the eigenfunctions $\phi_n(v)$ with eigenvalues $\lambda_n$ form a basis of the functional space of the probability densities $p$, the equivalence
\begin{equation}
\begin{split}
   p(v,t) = & \phi_0(v) + \sum_{n=1}^\infty a_n(t) \phi_n(v) \\
          \equiv & \phi_0(v) + \vec{\phi}(v) \cdot \vec{a}(t)
\end{split}
\label{eq:DensityDecomp}
\end{equation}
holds with $\lambda_0 = 0$ and $a_0 = 1$ \cite{Mattia2002}.

From this decomposition we note that the moments can be expressed by the linear combination
\begin{equation*}
   \vec{m} = \vec{m}_0 + \mathbf{U} \vec{a}
\end{equation*}
obtained by applying to both hand sides of the equation the bra $\bra{v^n}$, and singling out the moments computed on the stationary eigenmode (i.e., the probability density where the system relaxes under stationary condition): $\{\vec{m}_0\}_k = m_{k0} \equiv U_{k0}$.
Here, the matrix $\mathbf{U}$ has elements
\begin{equation}
    U_{kn} \equiv \braket{v^k}{\phi_n} = \int_\vmin^\vthr v^k \phi_n(v) \, ,
\label{eq:Umatrix}
\end{equation}
similar to the expected values of $v^k$ if $\phi_n(v)$ would be probability densities, which is not the case.
Under the hypothesis that $\mathbf{U}$ is invertible, an intriguing expression for the coefficients of the spectral decomposition results
\begin{equation}
	\vec{a}=\mathbf{U^{-1}}(\vec{m}-\vec{m}_0) \, .
\label{eq:MDecomp}
\end{equation}
Once inserted in Eqs.~\eqref{eq:MDynStart} and \eqref{eq:DensityDecomp} a self-consistent moment dynamics is obtained for an uncoupled set of spiking neurons (i.e., $J = 0$). 
This infinite set of ordinary differential equations represents an alternative decomposition to the one describing the evolution in time of the expansion coefficients $\vec{a}$ \cite{Mattia2002}, and as such, fully equivalent to the FP equation \eqref{eq:FP}.

When recurrent synaptic interactions are considered (i.e., $J \neq 0$), the infinitesimal mean and variance of the input current in \EqRef{eq:MuSigma} explicitly depends on the firing rate $\nu$. 
Self-consistency then requires to express also this state variable in term of $\vec{m}$.
This is a straightforward task as from \EqRef{eq:DensityDecomp}, $\nu$ is simply given by a linear combination of the expansion coefficients \cite{Mattia2002}:
\begin{displaymath}
	\nu = \Phi(\mu,\sigma) + \vec{f} \cdot \vec{a} \, .
\end{displaymath}
Here $\Phi(\mu,\sigma) = S_{\phi_0}(\vthr)$ is the so-called current-to-rate gain function returning the firing rate under stationary conditions, while the flux vector of non-stationary modes can be conveniently written as $\vec{f} = \vec{1}/\tau$ \cite{Deniz2016}. 
Now, using the $\vec{m}$-dependency of $\vec{a}$ in \EqRef{eq:MDecomp}, the firing rate can be eventually rewritten as 
\begin{equation}
   \nu(t) = \Phi(\mu,\sigma) +\frac{1}{\tau} \vec{1} \cdot \mathbf{U}^{-1}(\vec{m}-\vec{m_0})
   \label{eq:NuVsMoments}
\end{equation}
This expression is general being valid for any IF neuron model provided that the proper $\Phi$ and $\mathbf{U}$ are used. 
Note that the relation between firing rate and moments of membrane potential is deceptively linear. 
In fact, all the terms depend on both $\mu$ and $\sigma$, function themselves of the firing rate, and thus of the density $p(v,t)$.

Even in this mean-field framework, Eqs. \eqref{eq:MDynStart}, \eqref{eq:DensityDecomp},
% \eqref{eq:MDynVec}, 
\eqref{eq:MDecomp} and \eqref{eq:NuVsMoments} describe the self-consistent dynamics of a population of interacting spiking neurons, and are fully equivalent to FP equation \eqref{eq:FP} when the current moments in \EqRef{eq:MuSigma} are taken into account.
In principle, such decomposition in infinite moments does not add any advantage with respect to the original spectral decomposition \EqRef{eq:SpecDecomp}, as it is needed to compute the matrix $\mathbf{U}$.
However, as we show in the following, there are specific problems where a recursive expressions for the moments and the elements of $\mathbf{U}$ exists allowing to consider only a limited number of degrees of freedom.

\paragraph{Network of leaky integrate-and-fire neurons ---}
Focusing on the well-known `leaky' integrate-and-fire (LIF) neuron, the relaxation drift is $F(V) = -V/\tau$ with decay time $\tau$, and the dynamics \eqref{eq:MDynStart} can be explicitly worked out for any $k \geq 1$ \cite{Ricciardi1977,Shao2020}:
\begin{equation}
   \frac{\dot{m}_k}{k} = - \frac{m_k}{\tau} + \mu m_{k-1} + \frac{k-1}{2}\sigma^2 m_{k-2}  - \frac{\beta_k}{k} \nu \, ,
\label{eq:RecurMomDyn}
\end{equation}
where $\beta_k = v_\mathrm{thr}^k - v_\mathrm{res}^k$, and $m_0=1$.

As expected from having explicitly introduced the flux reinjection \eqref{eq:ReinjBC}, the moment dynamics is explicitly driven by the population firing rate $\nu(t)$.
This requires to solve \EqRef{eq:NuVsMoments}, and hence to derive an explicit expression for the matrix $\mathbf{U}$.
Analogously to the moments of the membrane potential, by applying the operator $\int_\vmin^\vthr v^n$ to both hand sides of \EqRef{eq:SpecDecomp} we obtain
%\begin{equation*}
%    U_{kn}= \langle v^k \rangle_{\phi_n} = \frac{-\beta_k +k\mu U_{k-1,n}  +\frac{\sigma^2}{2}k(k-1) U_{k-2,n}   }{\lambda_{n}\tau + k}
%\end{equation*}
\begin{displaymath}
    \left(\lambda_{n} + \frac{k}{\tau}\right) U_{kn} = k \mu U_{k-1,n}  +\frac{k(k-1)}{2} \sigma^2 U_{k-2,n} -\frac{\beta_k}{\tau} \, .
\end{displaymath}
%and by recursion a closed form can be found.
A recursive expression then arises
\begin{equation}
     U_{kn} = \frac{k \mu \tau U_{k-1,n} + k(k-1) \sigma^2 \tau U_{k-2,n}/2 - \beta_k}{\lambda_n \tau + k} \, ,
\label{eq:ULIF}
\end{equation}
where we must take into account that $U_{0n} = 0$, $U_{1n} = -\beta_1/(1+\lambda_n \tau)$ for any non-stationary mode ($n\neq 0$).

\paragraph{Low-dimensional population dynamics ---}
The above recursive expressions for networks of LIF neurons have a peculiar feature: the generic $k$-th order expression depends only on those at lower orders.
A hierarchy of moments and matrix $\bf{U}$ arises analogous the one for the eigenvalues $\lambda_n$.
This allows to truncate the decomposition to the first eigenmodes/moments. 
For instance, focusing on the first two moments we obtain 
\begin{equation}
\begin{split}
   \dot{m}_1 &= - \frac{m_1}{\tau} + \mu - \beta_1 \nu \\
   \dot{m}_2 &= - \frac{2}{\tau} m_2 + 2 \mu m_1 + \sigma^2 - \beta_2 \nu \, ,
\end{split}
\label{eq:2DMomentsDyn}
\end{equation}
from which the evolution of the variance $s = m_2 - m_1^2$ can be easily worked out:
\begin{equation}
   \dot{s} = - \frac{2 s}{\tau} + \sigma^2 - \left(\beta_2 - 2 \beta_1 m_1\right) \nu \, .
\label{eq:VarianceDyn}
\end{equation}

Now, to close the loop we need to derive the firing rate from \EqRef{eq:NuVsMoments}, which given the above $U_{1n}$ and
\begin{displaymath}
   U_{2n} = \frac{2 \mu \tau U_{1n} - \beta_2}{\lambda_n \tau + 2} \, ,
\end{displaymath}
after some algebra results to be
\begin{equation}
   \nu = \Phi + G_1(\lambda_{1,2}) (m_1 - m_{10}) + G_2(\lambda_{1,2}) (m_2 - m_{20})
\label{eq:2DFiringRate}
\end{equation}
with 
\begin{displaymath}
\begin{split}
   G_1(\lambda_{1,2}) & = - \frac{2 \mu \tau  
           [(\lambda_1+\lambda_2) \tau + 3] + \frac{\beta_2}{\beta_1}
   (\lambda_1 \tau + 1) (\lambda_2 \tau + 1)}{(2 \beta_1 \mu \tau - \beta_2)\tau} \\ 
   G_2(\lambda_{1,2}) & = \frac{(\lambda_1 \tau + 2) (\lambda_2 \tau +2)}{(2 \beta_1 \mu \tau - \beta_2) \tau} \, ,
\end{split}
\end{displaymath}
and
\begin{displaymath}
\begin{split}
   m_{10} & = -\beta_1 \Phi \, \tau \\
   m_{20} & = \mu \tau m_{10} - \frac{\beta_2}{2}
            = -\beta_1 \mu \tau^2 \Phi - \frac{\beta_2}{2}
\end{split} \, .
\end{displaymath}
In the limit of a network dynamics where the impact of higher-order nonstationary modes (i.e., $k,n>2$) is negligible, Eqs. \eqref{eq:2DMomentsDyn} and \eqref{eq:2DFiringRate} are expected to accurately predict the evolution of the first two moments of the membrane potential together with the firing rate of the whole population.

% \mmnote{HERE\dots}

\begin{figure}[!htp]
    \includegraphics[width=78mm]{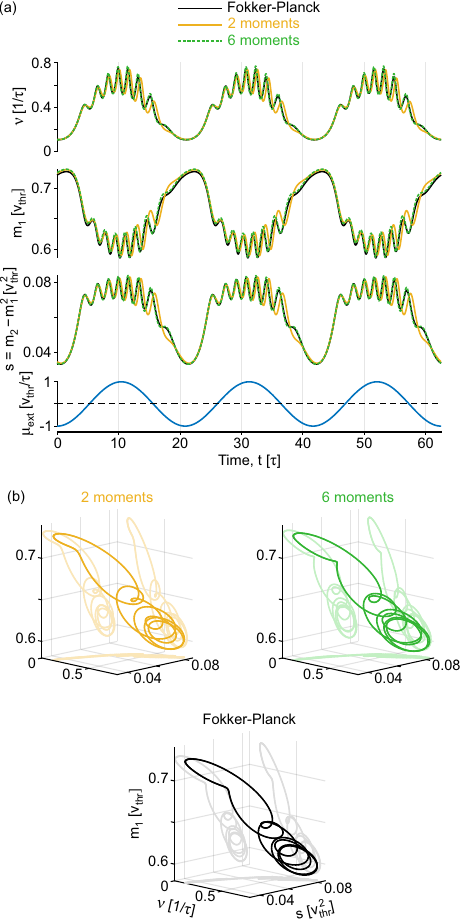}
    \caption{Moment decomposition effectiveness in a strongly nonlinear dynamical regime. 
    A network of excitatory LIF neurons is set to have $K=1000$, $J=0.38\vthr/K$ and $\mu_\mathrm{ext}$, $\sigma_\mathrm{ext}$ such that the stationary state is at firing rate $\nu_0 =0.4/\tau$ ($\mu_0=1.05 \vthr/\tau$, $\sigma_0 = 0.133 \vthr/\sqrt{\tau}$ and $\tau$=1). 
   With these parameters the fixed point is unstable and the rate dynamics is attracted to a limit cycle \cite{Vinci2023}.
   Each neuron of the network receives an additional external current $I_1 \sin(\omega t)$ ($I_1=0.12 \mu_\mathrm{ext}$, $\omega / 2 \pi \nu_0 = 0.12$). 
   The match between the trajectories of the firing rate and the first two moments is reported in panel (a) together with the related phase portrait in (b). 
   Orange, 2 moments system from Eqs. \eqref{eq:2DMomentsDyn} and \eqref{eq:2DFiringRate}; green, 6 moments systems from Eqs. \eqref{eq:MDecomp} and \eqref{eq:NuVsMoments}; black FP \EqRef{eq:FP}.}
    \label{fig:NLDYN}
\end{figure}

\paragraph{Moment decomposition effectiveness ---}
A great advantage of the presented theory is that it is valid even in strongly nonlinear regime and in open systems i.e, with external sources. 
We demonstrate this investigating the case of a strongly coupled network of excitatory neurons. 
Using the prediction of linear theory we chose the value of the synaptic strength $J$ such that the stationary solution is unstable and the firing rate dynamics follows a limit cycle. 
Moreover we add an external sinusoidal modulation of the mean synaptic current $\mu_0(t)$. 
As shown in \FigRef{fig:NLDYN} two moments alone can already faithfully reproduce most of the features of the dynamics obtained by direct integration of the FP \EqRef{eq:FP}. 
The quality of the match increase when we consider the decomposition including the first 6 moments, highlighting the hierarchical nature recursive expressions derived above. 

\begin{figure}
   \includegraphics[width=74mm]{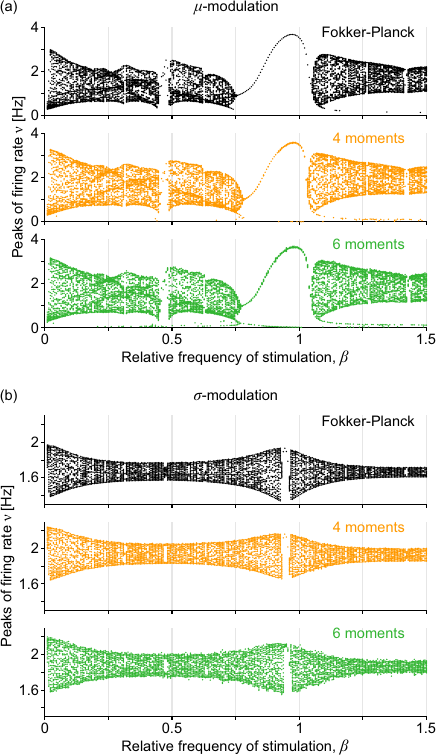}
   \caption{Entrained and aperiodic responses are fully capured by by low-dimensional moment dynamics.
   Points represent the singled-out local maxima if the firing rate of the 4 moments (orange) and 6 moments (green) systems and of the FP equation (black). 
   (a) sinusoidal modulation of the mean input current ($\mu(t) =KJ\nu(t) + 0.05 \mu_\mathrm{ext} \sin(2\pi \beta f_\mathrm{LC} t)$).
   (b) same sinusoidal stimulation affecting only the variance of the input current ($\sigma^2(t) =K J^2 \nu(t) + \sigma_\mathrm{ext}^2 + 0.03 \sigma_\mathrm{ext}^2 \sin(2 \pi \beta f_\mathrm{LC} t)$). 
   The stationary moments are $\mu_0=1.36 \vthr$ and $\sigma_0 =0.205 \vthr$ with $J=0.52\vthr/K$.
   The other network parameters are as in \FigRef{fig:NLDYN}.}
\label{fig:PeakMaps}
\end{figure}

Under this dynamical regime, the neuronal population exhibits an intrinsic frequency determined by its limit cycle. 
By varying the relative frequency of the sinusoidal perturbation, defined as $\beta = \omega / (2\pi f_\mathrm{LC})$ (where $f_\mathrm{LC}$ is the oscillation frequency of the spontaneous limit cycle), one can expect the emergence of aperiodicity and entrainment.
We then singled out the local maxima of the firing rate as a function of the stimulation frequency. 
In \FigRef{fig:PeakMaps} the distribution of such $\nu$-peaks are shown for modulations of both the mean $\mu_\mathrm{ext}$ and the variance $\sigma^2_\mathrm{ext}$ of external current varying $\beta$. 
Crowded regions signal complex structure of the firing rate dynamics with multiple local peaks possibly resulting from an aperiodic or chaotic-like dynamics.
Remarkably, in both kind of modulations, the aperiodic/chaotic regimes (i.e., the one with wide distributions of peaks) and entrainment when few different peaks are singled out, occur in the same range of relative stimulation frequency $\beta$.
Also the amplitude of these distribution faithfully reproduce the reference result coming from the numerical integration of the FP equation.

\paragraph{Discussion ---}
In this work have shown how it is possible to derive a self-consistent dynamics for the moments of the membrane potential distribution of a population of spiking neurons and, most importantly, their relation with the population firing rate.
Of particular interest in this regards are the eigenvalues of the FP operator that governs the mean-field population dynamics that have central roles and can be related to the moment of the single neuron inter-spike intervals \cite{Schaffer2013, Pietras2020, Vinci2024}. 
The dynamical system we derived has the great advantage to be valid both in noise and drift-dominated regimes in contrast with most moment closure approaches that can be found in the literature so far that only applies in restricted regimes. 
Those restrictions arises from approximations necessary to close the moment systems, such as assuming deterministic synaptic current \cite{Montbrio2015}, weak coupling \cite{Schaffer2013} or adiabatic approximation \cite{zhang2015, Shao2020}. 
%We could avoid these typical issue applying spectral expansion tools \cite{VinciTeoria}. 
As excepted the quality of reproduction of the macroscopic quantities increase with number of moments used which in our context is equivalent to the number of eigenmodes, or time scales, considered.
From this we derived specific (and explicit) analytic expressions for networks of LIF neurons.

Our theory works equally well in strongly nonlinear regimes both due to huge coupling of the network or strong external modulation.
By investigating the nonlinear response to exogenous stimulation, we found that the decomposition including only the first 2 moments  can already reproduce faithfully the firing rate dynamics.
We then considered a sinusoidal forcing of the mean $\mu$ or the variance $\sigma^2$ of the input current in the network with strong couplings $K J$ autonomously oscillating.
Also in this case the rich patterns of local maxima in $\nu(t)$ varying the frequency of the added oscillatory input, are fully reproduced when only a limited number of moments is taken into account.

Contrary to standard moment-closure approaches where an ansatz of $p(v,t)$ is assumed and consequently the macroscopic equation are consistently derived, we relied on the separation of time scales that is manifested in the spectral gap typical of spiking neural networks population dynamics \cite{Mattia2002, Vinci2024}. 
Our approach is thus expected to have a wider applicability.
For this reason we believe may allow to strengthen the link between theory and experiments.

\paragraph{Acknowledgments. ---}

Work partially funded by the NextGenerationEU and MUR (PNRR-M4C2I1.3) project MNESYS (PE0000006–DD 1553 11.10.2022) and project EBRAINS-Italy (IR0000011– DD 101 16.6.2022) to MM.

\bibliographystyle{apsrev4-2}
\bibliography{RefLibrary}% Produces the bibliography via BibTeX.

\end{document}